%% file: wwxsec189_final.tex
\documentclass[12pt]{article}
\usepackage{a4,epsfig,amssymb}
\def\PL#1#2#3{{Phys. Lett. }{\bf B#1 }(#2) #3}
\def\NIM#1#2#3{{Nucl. Inst. Meth. }{\bf #1 }(#2) #3}
\def\ra{\rightarrow}
\newcommand{\PW}{\mbox{$\mathrm{W}$}}

\newcommand{\PZz}{\mbox{$\mathrm{Z}$}}
\newcommand{\PMW}{\mbox{$m_{\mathrm{W}}$}}
\newcommand{\PMMW}{\mbox{$m_{\mathrm{W}}^2$}}
\newcommand{\PMMZ}{\mbox{$m_{\mathrm{Z}}^2$}}
\newcommand{\PSW}{\mbox{$\sigma_{\mathrm{WW}}$}}

\newcommand{\GEV}{\mbox{$\mathrm{GeV}$}}

\newcommand{\GEVcc}{\mbox{$\mathrm{GeV}/{{\it c}^2}$}}

\newcommand{\Pepem}{\mbox{$\mathrm{e}^+\mathrm{e}^-$}}
\newcommand{\Pee}{\mbox{$\mathrm{ee}$}}
\newcommand{\PWpWm}{\mbox{$\mathrm{W}^+\mathrm{W}^-$}}
\newcommand{\PWW}{\mbox{$\mathrm{W}\mathrm{W}$}}
\newcommand{\PZZ}{\mbox{$\mathrm{Z}\mathrm{Z}$}}

\newcommand{\ckm}[1]{\ensuremath{V_\mathrm{#1}}}
\newcommand{\ev}{\mbox{$\mathrm{e}\nu$}}
\newcommand{\mv}{\mbox{$\mu\nu$}}
\newcommand{\tv}{\mbox{$\tau\nu$}}
\newcommand{\lv}{\mbox{$\ell\nu$}}

\newcommand{\qqbar}{\mbox{$\mathrm{q\bar{q}}$}}
\newcommand{\stat}{\ensuremath{\mathrm { (stat.)}}}
\newcommand{\syst}{\ensuremath{\mathrm { (syst.)}}}
\newcommand{\pb}{\ensuremath{\mathrm {pb}}}
\newcommand{\ipb}{\ensuremath{\mathrm {pb}^{-1}}}

\begin{document}

\begin{titlepage}

{\large\centerline{EUROPEAN ORGANISATION FOR NUCLEAR RESEARCH (CERN)}}
\vspace{1cm}
\begin{flushright}
\begin{tabular}{rr}
         & CERN-EP/2000-052 \\
          & 11 April 2000 \\
\end{tabular}
\end{flushright}

\begin{center}
\vspace{3cm}
\boldmath
{\huge\bf Measurement of W-pair
production in  \Pepem{} collisions
at~189 GeV}\\
{\Large \vspace{7ex} The ALEPH Collaboration}
\unboldmath
\end{center}

\vspace{3cm}
\begin{center}
{\bf Abstract}
\end{center}
The production of $\PWpWm$ pairs is analysed in a data sample
collected by ALEPH at a mean centre-of-mass energy of 188.6~\GEV,
corresponding to an integrated luminosity of 174.2~$\ipb$. Cross
sections are given for different topologies of W decays into
leptons or hadrons. Combining all final states and assuming
Standard Model branching fractions, the total W-pair cross section
is measured to be $ 15.71 \pm 0.34 \stat \pm 0.18 \syst ~\pb$.
Using also the W-pair data samples collected by ALEPH at lower
centre-of-mass energies, the decay branching fraction of the W boson
into hadrons is measured to be~$\mathrm{B (W\ra hadrons)} = 66.97
\pm 0.65 \stat \pm 0.32 \syst\%$, allowing a determination of the
CKM matrix element~$|\ckm{cs}|= 0.951 \pm 0.030 \stat \pm 0.015
\syst$.
\\[1.5cm]

\vspace{1cm}
\centerline{\em (Submitted to Physics Letters B)}
\end{titlepage}


\input{authb}

\section{Introduction}
This letter presents results on  W-pair production in \Pepem collisions
using data collected with the ALEPH detector at a
centre-of-mass (CM) energy around 189~\GEV, during the 1998 data taking period.
The \PWW\ events are identified in all possible W decay channels,
thus allowing the determination of the W branching fractions 
and indirectly, the coupling of the W to cs pairs.

The experimental conditions and data analysis follow
those used in the cross section measurements at lower LEP2 energies.
As they are already described in detail in~\cite{xsecA},
attention is focused here on changes to selection procedures
other than a simple rescaling of cuts with the increased collision energy.



A detailed description of the ALEPH detector can be found in
Ref.~\cite{det} and of its performance in Ref.~\cite{perf}. The
luminosity is measured from small-angle Bhabha events, using
lead-proportional wire sampling calorimeters~\cite{ALEPHlcal},
with an accepted Bhabha cross section of approximately
4.25~nb~\cite{BHLUMI}. An integrated luminosity of
174.20~$\pm$~0.20~$\stat~\pm$~0.73~$\syst~\ipb$ was recorded at a
mean CM energy of
188.63~$\pm$~0.04~\GEV~\cite{LEPbeam}.

In this letter, the quoted signal cross sections are the CC03 
cross sections~\cite{LEP2workshop}, 
defined as the production of four-fermion final states through two
resonating W bosons.  Two processes contribute, 
$\nu_{\mathrm{e}}$ exchange in the $t$-channel and
$\mathrm{Z/}\gamma$ exchange in the $s$-channel.
The measured cross sections
are corrected for the difference, denoted the ``4f-CC03
correction''~\cite{xsecA}, between the accepted cross sections for CC03 
processes and all Standard Model four-fermion final states consistent with
W-pair decays.

The CC03 Standard Model cross section (\PSW), calculated at 
$\sqrt{s}= 188.63$ GeV with the program~{\tt GENTLE}~\cite{GENTLE} is 
16.65 pb ($\pm 2\%$). The {\tt KORALW}~\cite{KORALW} version 
1.21 Monte Carlo event generator is used to simulate the signal events with 
a normalised cross section in agreement with the {\tt GENTLE} value. 
The {\tt JETSET}~\cite{Jetset} package is used for the hadronisation.
Comparison samples, generated with {\tt
EXCALIBUR}~\cite{EXCALIBUR} and {\tt grc4f}~\cite{grc4f} for both 
CC03 and all four-fermion diagrams, are used for systematic error
evaluation. Samples of events are also generated with different W
masses, both for CC03 diagrams and for all WW-like four-fermion
diagrams, with {\tt KORALW}.


The {\tt KORALZ}~\cite{KORALZ} Monte Carlo program
is used to generate $\Pepem\rightarrow\qqbar$ background events.
Other backgrounds are generated with {\tt PYTHIA 5.7}~\cite{PYTHIA} for
\PZZ, \PZz\Pee ~and \PW\ev ~processes, {\tt PHOT02}~\cite{PHOT02}
for $\gamma\gamma$ interactions, {\tt KORALZ} for $\mu$ and $\tau$
pair production and {\tt BHWIDE}~\cite{BHWIDE} and {\tt
UNIBAB}~\cite{UNIBAB} for Bhabha events.

\boldmath
\section{Selection of W-pair candidates}
\subsection{$ \PWW \ra \ell\nu \ell\nu$ events}
\unboldmath

The selection of fully leptonic W-pair decays follows exactly
the two selections used for the cross section and
branching fraction measurements at 183~GeV~\cite{xsecA}. The two selections 
have similar overall efficiencies and background levels but differ in their 
sensitivities to the individual dilepton channels. The first is based
on topological information and is sensitive to all channels. In the second, 
lepton identification is used to optimise the cuts according to which final 
state is being considered. Events are accepted as \PWW\ candidates if they 
pass either of the two selections and are then classified into six di-lepton 
channels making use of electron and muon identification criteria. A jet or a 
single charged particle track is classified as a tau if no lepton is 
identified or the identified lepton has an energy less than 25~GeV. 

Beam related background, not simulated
in Monte Carlo events, affects the efficiency of the cut which
removes events depositing energy within 
$12^{\circ}$ of the beam. Random trigger events were used to 
model these local energy deposits close to the beam. The 
inefficiency introduced is found to be $8.2 \pm 0.5 \%$. 
The CC03 efficiencies in the individual $\ell\nu\ell\nu$ channels
are given in Table~\ref{tab:summ} after correction for this beam
related effect. The inclusive combination of the two selections
has an overall efficiency of $64.2 \pm 0.4\%$ for the fully leptonic channels 
when combined assuming lepton universality. The total
background amounts to $131 \pm 7 \stat \pm 8.5 \syst$ fb and
is dominated by $\gamma\gamma \ra \ell\ell$ and non-WW-like
$\mathrm{ZZ} \ra \ell\ell\nu\nu$ events.
In the data, the inclusive combination selects 220 events.

All sources of systematic uncertainties are listed in
Table~\ref{tabsys}. They are dominated by 
the uncertainties on the background cross sections, the
uncertainty from the cut on energy detected close to the
beam and by Monte Carlo statistics.

A maximum likelihood fit is applied to determine the cross
section for each fully leptonic decay channel using the efficiency
matrices for signal and backgrounds given in Table~\ref{tab:summ}.
For all channels together the 4f-CC03 correction is $-10\pm 10$ fb,
where the uncertainty comes from Monte Carlo statistics.

The results of the fit are
\begin{eqnarray*}
\sigma(\PWW\ra \ev\ev ) & = &
0.19 \pm 0.05 \stat \pm 0.01 \syst ~\pb, \\
\sigma(\PWW\ra \mv\mv ) & = &
0.20 \pm 0.05 \stat \pm 0.01 \syst ~\pb, \\
\sigma(\PWW\ra \tv\tv ) & = &
0.22 \pm 0.08 \stat \pm 0.02 \syst ~\pb, \\
\sigma(\PWW\ra \ev\mv ) & = &
0.43 \pm 0.07 \stat \pm 0.01 \syst ~\pb, \\
\sigma(\PWW\ra \ev\tv ) & = &
0.36 \pm 0.08 \stat \pm 0.02 \syst ~\pb, \\
\sigma(\PWW\ra \mv\tv ) & = &
0.38 \pm 0.08 \stat \pm 0.01 \syst ~\pb.
\end{eqnarray*}
The systematic uncertainties are obtained by varying all input
parameters in the fit
according to their uncertainties.

The total fully leptonic cross section is obtained with the same
fit, assuming lepton universality: $$ \sigma (\PWW\ra
\ell\nu\ell\nu) = 1.78 \pm 0.13 \stat \pm 0.02 \syst ~\pb, $$ 
consistent with the sum of the individual channels.


\renewcommand{\arraystretch}{1.2}
\begin{table}[h]
 \begin{center}
\caption{\protect\footnotesize Summary of results of the different event 
selections on Monte Carlo and data events. Efficiencies are given in percent of
  CC03~processes; they assume Standard Model branching fractions in the 
overall values quoted in the right-hand column.
The \qqbar\qqbar~column refers to events with a NN output greater than 0.3 
where the backgrounds also include non-\qqbar\qqbar\ \PWW\ decays.
  The listed backgrounds do not include the 4f-CC03 corrections.
} \vspace{6pt} \label{tab:summ}
 \begin{tabular}{|c|l|cccccc|ccc|c||c|} \hline
\multicolumn{2}{|c|}{} &\multicolumn{11}{c|}{Event selection and
classification}    \\
 \cline{3-13}
\multicolumn{2}{|c|}{} &
 $ee$&$e\mu$&$e\tau$&$\mu\mu$&$\mu\tau$&
 $\tau\tau$&$e\qqbar$&$\mu\qqbar$&$\tau\qqbar$&$\qqbar\qqbar$&
 All\\
 \hline
& $\ev\ev\!\!$  & 57.8 &   -  &  8.8 &   -  &   -  &  0.5
                &  -   &   -  &  -   &   -  & 67.1 \\
& $\ev\mv\!\!$  &   -  & 59.0 &  4.7 &   -  &  4.6 &  0.3
                &  -   &   -  &  -   &   -  & 68.6 \\
& $\ev\tv\!\!$  &  3.0 &  4.2 & 50.1 &   -  &  0.3 &  4.3
                &  -  &   -  &   -   &   -  & 61.9  \\
Eff. for & $\mv\mv\!\!$  &   -  &  - &   -  & 61.9 &  8.3 &  0.3
                &   -  &   -  &   -  &   -  & 70.6 \\
$\PWW\!\!\ra$ & $\mv\tv\!\!$  &   -  &  4.2 &  0.3 &  3.5 & 52.9 &
3.5
                &   -  &  -   &   -  &   -  & 64.4 \\
(\%) & $\tv\tv\!\!$  & 0.2  &  0.4 &  7.7 &  0.4 &  6.0 & 36.5
                &   -  &   -  &   -  &   -  & 51.2 \\
 \cline{2-13}
& $\ev\qqbar\!\!\!$&  -   &   -  &   -  &   -  &   -  &   -
                & 82.4 &   -  &  5.4 & 0.2  & 88.0 \\
& $\mv\qqbar\!\!\!$&  -   &   -  &   -  &   -  &   -  &   -
                &  -   & 87.5 &  4.1 & 0.2  & 91.9 \\
& $\tv\qqbar\!\!\!$&  -   &   -  &   -  &   -  &   -  &   -
                &  3.7 &  3.8 & 59.0 & 0.9  & 68.0 \\
 \cline{2-13}
& $\qqbar\qqbar\!\!\!\!$
                &   -  &   -  &   -  &   -  &   -  &   -
                &   -  &   -  &   -  & 91.7 & 91.7 \\
 \hline\hline
\multicolumn{2}{|c|}{\parbox{2.5cm}{Expected background events}} &
  3 & 1 & 7 &  2 &  3 & 6
& 17 & 5 & 33 & 323 & 400 \\
 \hline\hline
\multicolumn{2}{|c|}{\parbox[c]{2.5cm}{Observed Events}} &
   24 &  51 &  48 &  26 &  46 &  25 & 381 & 382 & 303 & 1435 & 2721 \\
 \hline
\end{tabular}
\end{center}
\end{table}
\renewcommand{\arraystretch}{1.0}

\boldmath
\subsection{$\PWW \ra \ell \nu \qqbar$ events}
\unboldmath

As for the lower energy measurements, three $\ell\nu\qqbar$
selection procedures are applied. One selection requires  an 
identified electron or muon. The other two are designed to select
$\tau\nu\qqbar$ events, based on global variables or topological
properties of the events.

The selection of $\mathrm{e}\nu \qqbar$ and $\mu\nu \qqbar$ events
has been modified with respect to the previous analysis~\cite{xsecA}
to take into account the greater initial boost of the W's. 
The preselection remains similar. It is based
on the total charged particle energy and multiplicity and a cut 
on the longitudinal momentum and visible energy to reject Z$\gamma$ 
events with an undetected photon. 
The selection of the lepton track relies on the fact that it is
in general more energetic and isolated than the charged
particles from the hadronic system.
Thus the candidate lepton is chosen as the charged track that maximises
$p_{\ell}^2 (1-\cos\theta_{\ell j})$, where $p_{\ell}$ is the
track momentum and $\theta_{\ell j}$ the angle of the track to the
closest of the jets clustered using the remaining reconstructed charged 
particles. For the jet clustering the {\tt DURHAM-P}~\cite{durham} algorithm 
with a $y_{\rm cut}$ of 0.0003 is used. 

The same electron or muon identification criteria as for the fully
leptonic channels are required for this lepton candidate track. However, no
cut is applied on the lepton energy, so that \tv\qqbar~ events where
the $\tau$ decays to a softer lepton (as $\tau\ra\ev\nu$ or
$\tau\ra\mv\nu$) are also selected by this analysis.

For electron candidates the lepton energy is corrected for
possible bremsstrahlung photons detected in the
electromagnetic calorimeter. The isolation of the lepton is 
defined as $\log(\tan\theta_C/2)+\log(\tan\theta_F/2)$
where $\theta_C$ and $\theta_F$ are, respectively,
the angle of the lepton to the closest
charged track, and the opening angle of the largest cone centred
on the lepton direction which  contains a total energy smaller than
5 \GEV.

\begin{figure}[htb]
  \centerline{
    \epsfig{file=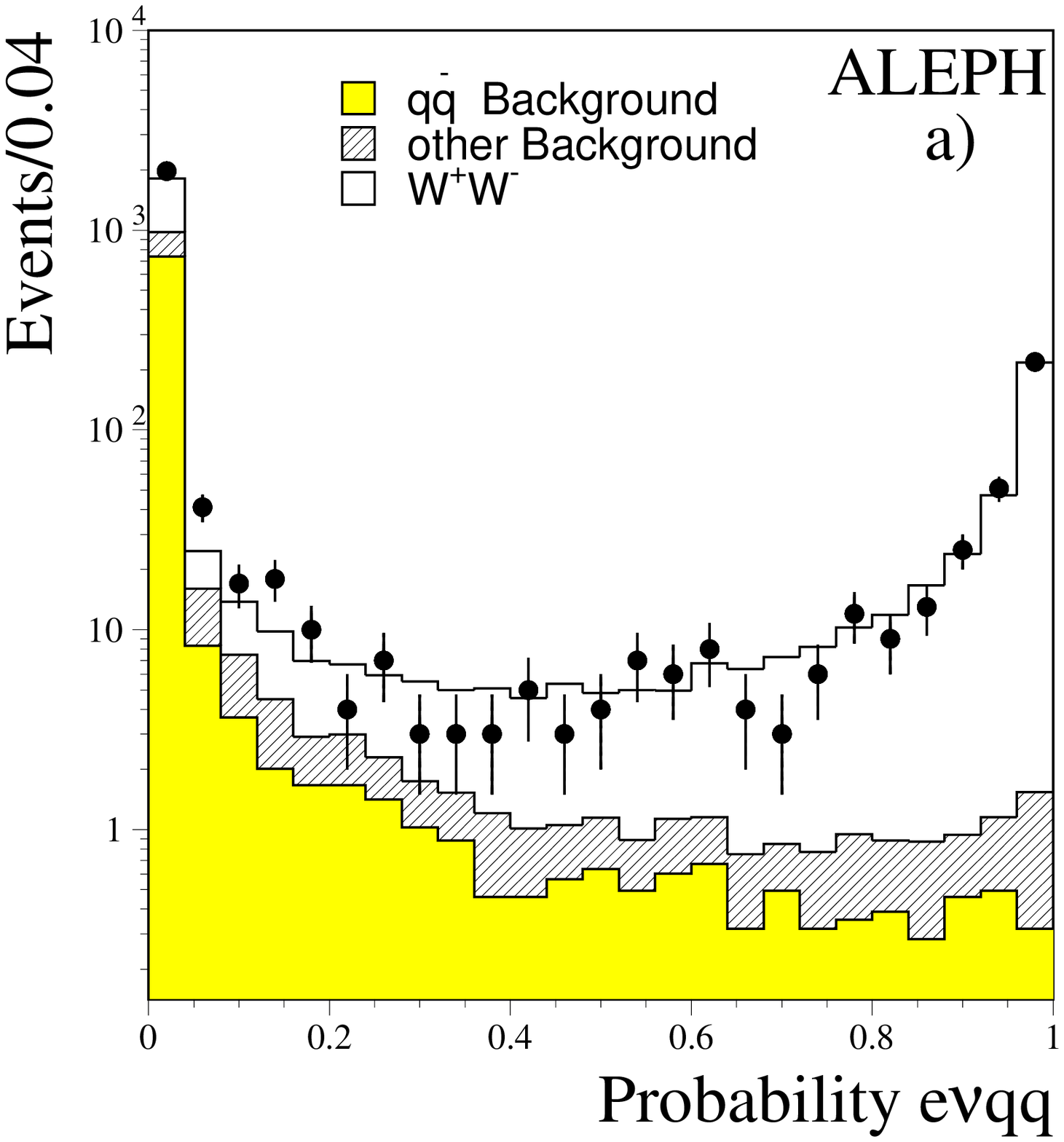,height=9cm}
    \epsfig{file=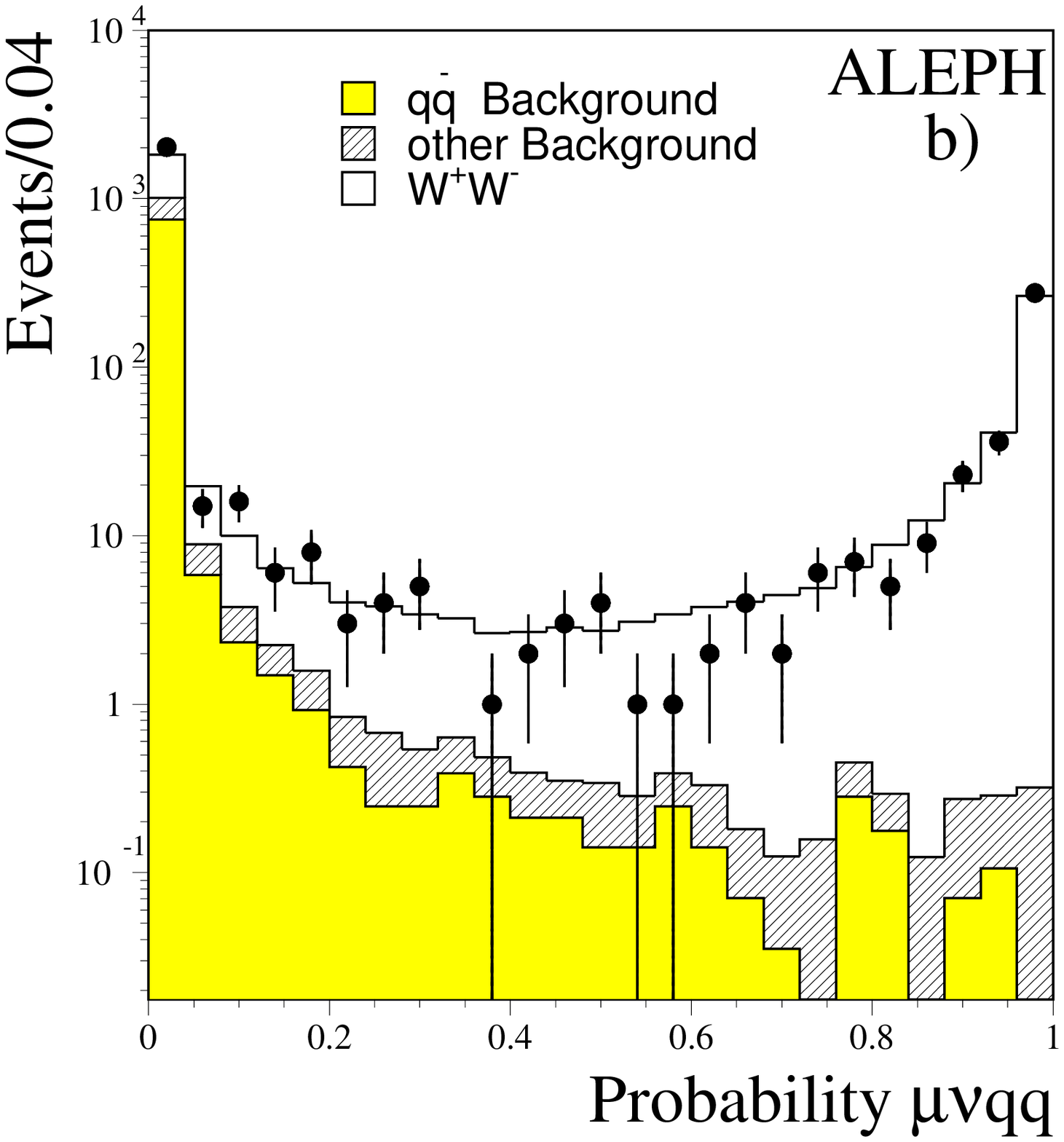,height=9cm}
    }
\hspace{3cm}
    \epsfig{file=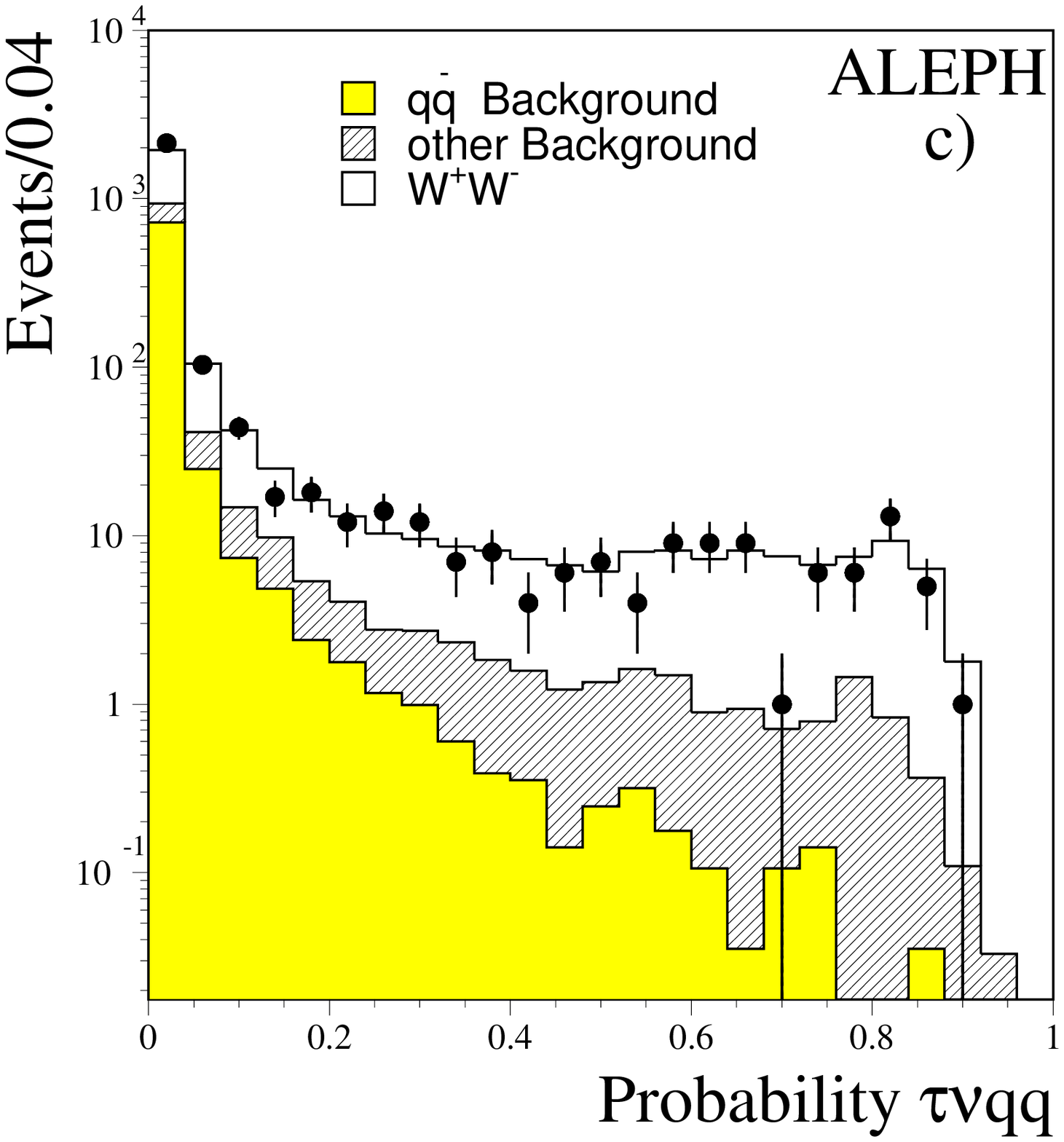,height=9cm}
\vspace*{-0.5cm}
 \caption{\protect\footnotesize Probability distributions of preselected
events for the
  (a) $\ev\qqbar$, (b) $\mu\nu \qqbar$ and (c) $\tau\nu \qqbar$ selections.
The points are the data and the histograms the Monte Carlo
predictions. The non-$\qqbar$ backgrounds include \PZZ, 
\PZz\Pee, \PW\ev\ and $\tau^+\tau^-$ processes.} 
\label{Plvqq}
\end{figure}

For each event, probabilities that it comes from each of the three 
signal processes, $\ev\qqbar$, $\mv\qqbar$ or $\tv\qqbar$, are determined 
(Fig.~\ref{Plvqq}) using Monte~Carlo reference samples of signal and 
backgrounds. These are evaluated from the identity, energy and isolation of 
the lepton plus the event total transverse momentum.
An event is classified as $\ev\qqbar$ or
$\mv\qqbar$ if its corresponding probability is greater than 0.40; it is then
not considered in the tau search.

The selection of $\tau\nu \qqbar$ events is based both on global event
variables and a topological selection which attempts
to identify the $\tau$ jet. As the selections were described in
detail in previous papers~\cite{xsecA}, only changes other
than a rescaling of the values of the energy-based cuts 
are described in the following:
\begin{itemize}
\item[(i)]
in the global analysis the visible mass is required to be greater
than 85~\GEVcc{} and less than 155~\GEVcc{} to take account of 
 the boost of the W boson.
The estimated energy of the ``primary'' neutrino must be smaller
than 70~\GEV;
\item[(ii)]
in the topological analysis, the energy of the most energetic
quark jet must be less than 75~GeV and the mass of the
hadronic system, i.e., the mass excluding the tau jet, is required
to be less than 100~\GEVcc{}.
\end{itemize}

In addition, if an event with a well defined e or $\mu$ fails 
the e/$\mu$ probability cut 
it is considered as a $\tau$ candidate and kept if the  $\tau\nu \qqbar$ 
probability (Fig.~\ref{Plvqq}c) is greater than~0.4. 

\begin{table}[htb]
\begin{center}
\caption[]{\protect\footnotesize Systematic error summary (units in fb)} 
\vspace{6pt}
\label{tabsys}
\begin{tabular}{|l|c|c|c|}
\cline{2-4} \multicolumn{1}{c|}{} & \multicolumn{3}{|c|}{\PWW\ cross
section}  \\ \hline Source &  $\ell\nu\ell\nu$ & $\ell\nu\qqbar $
&$\qqbar\qqbar$ \\ \hline Calibration of calorimeters  & & &  32 \\
\hline Jet calibrations & & & 9 \\ \hline \PWW\ generator and
$\PMW$ dependence & & & 25 \\ \hline \PWW\ fragmentation    & & &
17\\ \hline Lepton isolation & & 56 & \\ 
\hline \qqbar\ generator    & & & 23 \\ \hline Background shape
& & & 50\\ \hline Final state interactions & & & 38
\\ \hline Background normalisation & 13 & 30& 84 \\ \hline
Luminosity & 9 & 35 & 39 \\ \hline Monte Carlo statistics & 15 &
61 & 36
\\ \hline Beam related background & 11 & 30 & 40 \\ \hline Lepton
identification    & 2 & 45 & \\

\hline Probability cut & & 46 & \\ \hline
 \hline Total & 24 &  118 &  134
\\ \hline
\end{tabular}
\end{center}
\end{table}

\clearpage

\subsubsection{Results}
\label{reslvqq}

Table~\ref{tab:summ} gives the efficiencies for each selection.
The inclusive efficiencies for the three semileptonic decay
channels are $87.8 \pm 0.4$\% for the electron channel, $91.6 \pm
0.4$\% for the muon channel and $66.5 \pm 0.5$\% for the tau
channel, giving an $82 \pm 0.3$\% average efficiency for $\PWW\ra\lv\qqbar$, 
with a background of $314 \pm 12 \stat \pm 25 \syst$ fb.
The overall 4f-CC03 correction is
+8 $\pm$ 40 fb. A total of 1066 events are selected in the data.

The systematic uncertainties on the combined semileptonic cross
section  are summarised in Table~\ref{tabsys}. The largest contribution 
arises from the Monte Carlo statistical error on the 4f-CC03 correction.
Uncertainties arising from the choice of the lepton isolation criterium and
the probability cut are estimated from the change of efficiency following a
bin-by-bin reweighting of the respective Monte Carlo one-dimensional
distributions to the data.  Background normalisation mainly affects the
tau channel although there is also a contribution from residual Bhabha
background in the electron channel. Decays of the \PZz\ to electrons and
muons are used to evaluate lepton identification uncertainties whilst
the contribution from beam related backgrounds is estimated by
superimposing the energy deposits from random triggers on to the
simulated events.


To evaluate the individual cross sections a 
similar fit as for the fully leptonic events is used with the
corresponding matrix of efficiencies and backgrounds.
This yields 
\begin{eqnarray*}
\sigma (\PWW\ra \ev\qqbar) & = & 2.41 \pm 0.14 \stat \pm 0.07 \syst ~\pb, \\
\sigma  (\PWW\ra\mv\qqbar) & = & 2.39 \pm 0.13 \stat \pm 0.06 \syst ~\pb, \\
\sigma  (\PWW\ra\tv\qqbar) & = & 2.23 \pm 0.17 \stat \pm 0.08 \syst ~\pb,
\end{eqnarray*}
where the systematic uncertainties are obtained by varying all input
parameters of the fit according to their uncertainties.

The total $\lv\qqbar$ cross section is extracted by means of the same
fit, under the assumption of lepton universality:

$$ \sigma  (\PWW\ra\ell\nu \qqbar) = 7.07 \pm 0.23 \stat \pm 0.12
\syst~\pb,$$ 
again consistent with a simple sum of the three channels.

\boldmath
\subsection{$\PWW \ra \qqbar\qqbar$ events}
\unboldmath

The analysis of \PWW\ decays to four jets is updated from
Ref.~\cite{xsecA} and consists of a simple preselection followed
by a fit to the distribution of the output of a neural network
(NN) with 14 input variables.

In the preselection, the first step is to remove events with a large undetected
initial state (ISR) photon from radiative returns to the Z by 
requiring that the modulus of the total longitudinal momentum of all objects 
is less than $1.5 (M_{vis}-M_Z)$ where $M_{vis}$ is the observed visible mass.
The particles are then forced into four jets
using the {\tt DURHAM-PE} algorithm~\cite{durham} and the value of $y_{34}$, 
where a four-jet event becomes a three-jet event, is required to be greater 
than 0.001.
To reject $\qqbar$ events with
a visible ISR photon none of the four jets can have
more than 95\% of electromagnetic energy in a one degree cone
around any particle included in the jet. Four-fermion final
states where one of the fermions is a charged lepton are
rejected by requiring that the maximum energy fraction of a single
charged particle in a jet be smaller than 0.9.

\begin{figure}[htb]
\begin{center}
\mbox{\epsfig{file=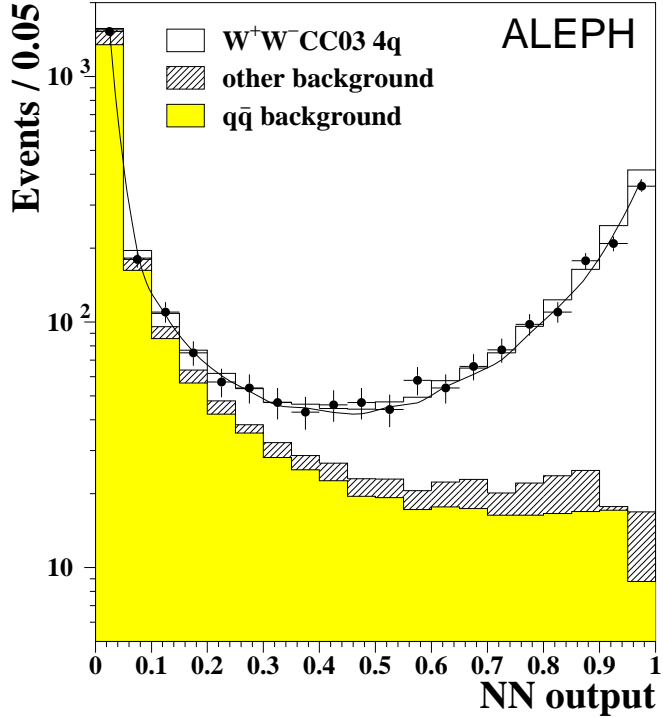,width=100mm}}
\end{center}
\vspace*{-0.1cm} 
\caption{\protect\footnotesize Comparison of NN output distributions
for data and Monte Carlo after the four-quark preselection. The
points are the data and the histograms the Monte Carlo
predictions. 
The line shows the fit result.
\label{had:nn}}
\end{figure}
At this point, 3438 events are selected in the
data while $3593 \pm 7$ events are expected from Standard Model
processes. This preselection has an efficiency of 98.4\% for CC03
events and a purity of 35.9\%. 

The input variables for the NN are described in the Appendix and
are related to the global event properties, the properties of
jets, \PWW\ kinematics and the b-tag probabilities for the four jets.
The NN is simplified with respect to that used for the analysis of
the 183 GeV data~\cite{xsecA}, with fewer input variables and a
slightly improved performance.
The output distributions of the NN for the data compared with the signal and 
backgrounds predicted by the Monte Carlo are given in Fig.~\ref{had:nn}. 
In Table~\ref{tab:summ} results are given for a cut at 0.3 on the NN 
value. 
 
The cross section is extracted 
by means of a binned maximum likelihood fit to the full NN output 
distribution. In the fit only the normalisation of the Monte Carlo signal 
is allowed to vary;
all backgrounds, including $\PWW \ra \ell \nu \qqbar$, are kept fixed both in 
shape and normalisation. Systematic studies at the Z peak when two jet 
$\qqbar$ events are forced into four jets show small discrepancies 
between data and Monte Carlo. This affects the $\qqbar$ background shape and 
is quantified using the 14 NN variables to yield a correction of $+50 \pm 50$ 
fb to the value from the fit.
The corrected fit result is: $$ \sigma (\PWW\ra
\qqbar\qqbar) = 6.89 \pm 0.23\stat \pm 0.13\syst~\pb. $$
\noindent
If only events with a NN output greater than 0.3 are kept a value
of $6.89 \pm 0.23~\pb$ is found showing no significant bias due to
the \qqbar\ background. 

The sources contributing to the systematic uncertainty are summarised in
Table~\ref{tabsys}. The largest contribution is the uncertainty in the \qqbar\ 
background normalisation where a 5\% variation is assumed.
The contributions from uncertainties in the values of jet variables  
used in the event preselection and the NN input are
assessed by adjusting their directions and energies according to
residual differences observed between Z calibration data and Monte 
Carlo~\cite{wmass_189}. Possible miscalibrations of the calorimeters are also 
taken into account. The beam background is studied in the same way as 
described in section~\ref{reslvqq}. The Monte Carlo statistics error is 
dominated by the 4f-CC03 correction.

Uncertainties in the \PWW\ generator are evaluated by comparing samples of 
fully simulated Monte Carlo events generated with {\tt KORALW} and 
{\tt EXCALIBUR}. To establish a 
\PWW\ fragmentation uncertainty, the {\tt HERWIG}~\cite{HERWIG} generator was 
tuned at the \PZz\ peak both for all flavours and non-b quark flavors.  
Then, the same samples of signal events generated with {\tt KORALW} are 
fragmented with both {\tt JETSET} and the appropriately tuned {\tt HERWIG}. 
For the 
\qqbar~background uncertainty, a sample of events generated with {\tt KORALZ} 
using {\tt JETSET} is compared with a separate sample of pure {\tt HERWIG} 
events to assess the effect of the choice of generator.  
Colour reconnection effects are estimated using the {\tt SK1} model in 
{\tt JETSET} with a reconnection probability of 0.3 and the effect of 
Bose-Einstein correlations is estimated according to the scheme, 
denoted $\rm{BE_3}$, as proposed for the {\tt LUBOEI}~\cite{BEJetset} 
implementation in {\tt JETSET}. 
The procedures followed are the same as those used in Ref.~\cite{wmass_189}.  


Several cross checks have been performed on the fit result to
search for possible biases arising in the selection and full
simulation of the \PWW\ events. The previous version of the analysis with a 
different preselection and NN~\cite{xsecA} gives $6.81
\pm 0.23~\pb$. Another estimate with a preselection based mostly
on charged tracks and a six variable NN using only charged tracks gives 
$6.83 \pm 0.25~\pb$. 
Also, a selection based purely on calorimeter measurements with six input
variables to a linear discriminant gives $6.91 \pm 0.26~\pb$. A
variety of linear discriminant analyses using from 4 to 14
variables give results which vary from 6.65 to 6.77 pb. All these checks give 
results which are consistently lower than the {\tt GENTLE} prediction.

\section{Total cross section}
The total cross section is obtained from a fit to all channels described above 
assuming the Standard Model branching fractions, the only unknown being the 
total cross section. The fit uses the matrices of
efficiencies and backgrounds for the various analyses and yields 
$$\PSW =  15.71 \pm 0.34\stat \pm 0.18\syst ~\pb. $$
\noindent Assuming no additional unexpected decay mode, the result is not 
significantly different if the branching fractions of the Standard Model 
decay modes are unconstrained.

The measurement is 5.5\% lower than the {\tt GENTLE} prediction. 
However, new calculations including full $\mathcal{O}\rm(\alpha)$ 
electroweak corrections, calculable in the double pole 
approximation (DPA)~\cite{Beenakker} have recently appeared. Two Monte Carlo 
programs, {\tt YFSWW3}~\cite{YFSWW} and {\tt RacoonWW}~\cite{RacoonWW}, are 
being developed. First numerical calculations find cross sections, 
respectively, 1.9\%~\cite{YFSWW} and 2.4\%~\cite{RacoonWW} lower than 
{\tt GENTLE} at 189 GeV.  
The predictions from {\tt RacoonWW} also include soft-photon exponentiation
and leading log corrections for initial state radiation beyond 
$\mathcal{O}\rm(\alpha)$ in addition to the calculations described 
in Ref.~\cite{RacoonWW}. 
The uncertainty in the two new models is expected to be of the order of  
0.5\%~\cite{MCLEP2}.

\begin{figure}[htb]
\begin{center}
\mbox{
 \epsfig{file=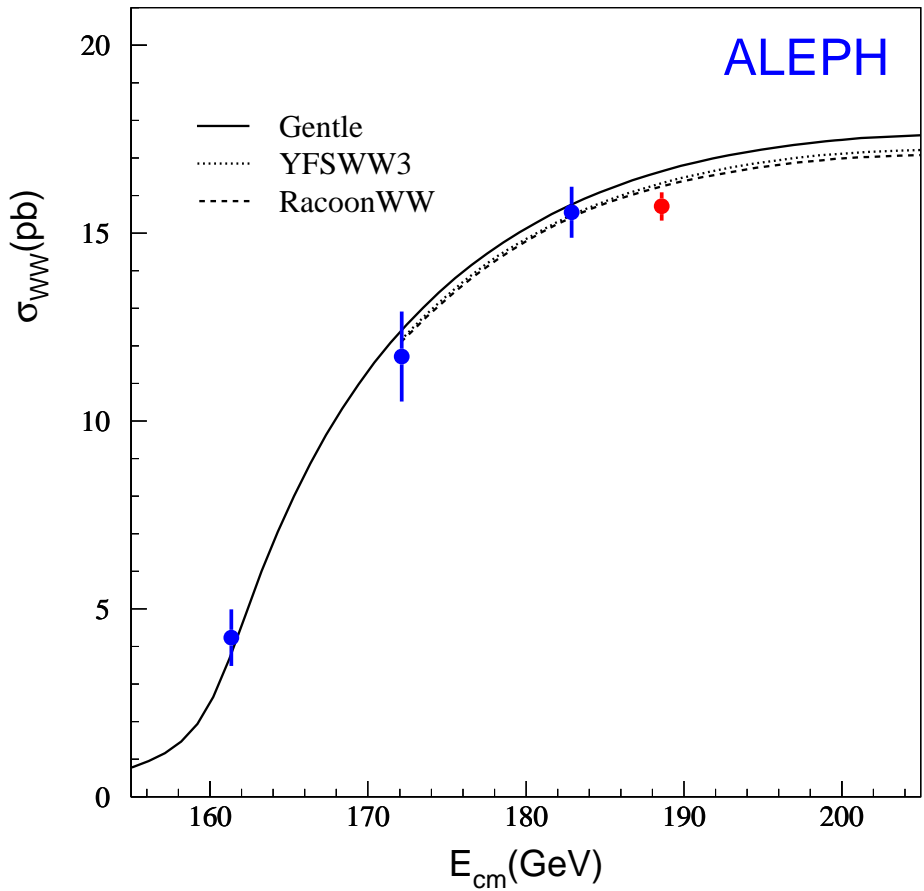,width=140mm}
}
\end{center}
\vspace*{-0.5cm} 
\caption{\protect\footnotesize Measurements of the W-pair production
cross section at four CM energies, compared with the
Standard Model predictions from {\tt GENTLE}, {\tt YFSWW3} and
{\tt RacoonWW}, for the LEP average value of the W
mass~\cite{MWWA}.
\label{fig:wwxs}}
\end{figure}

Fig.~\ref{fig:wwxs}
shows the total cross section measured as a function of the
CM energy. The predictions of the two more complete 
{\tt YFSWW3} and  {\tt RacoonWW} calculations are also shown and are 
in better agreement with the experimental results than {\tt GENTLE}. 
At 189~\GEV, the measurement is 3.8\% (1.5 standard deviations) below the 
{\tt YFSWW} prediction and 3.2\% (1.3 standard deviations) below the 
{\tt RacoonWW} prediction. Taking into account also the cross section values 
measured by ALEPH~\cite{xsecA} at 172 and 183~\GEV, the data are 
$4.6 \pm 2.0 \%, 2.8 \pm 2.0 \%$, and $2.3 \pm 2.0 \%$ below 
the {\tt GENTLE}, {\tt YFSWW} and {\tt RacoonWW} predictions respectively.
These values use the signal efficiencies determined with {\tt KORALW} and the 
quoted systematic uncertainty takes no account of any efficiency difference which 
may arise from the new calculations.

\section{Branching fractions and $V_{\rm cs}$}
\label{branchvcs} The same fit as for the total cross section is
performed combining the data samples collected at 161, 172, 183
and 189~\GEV~CM energies.

Without assuming lepton coupling universality, the seven
unknowns are the three individual leptonic branching fractions
and the four total
cross sections at 161, 172, 183 and 189~\GEV.
The hadronic branching fraction is set to
$1 - {\rm B}_{\mathrm e} - {\rm B}_{\mu} - {\rm B}_{\tau}$.
The fitted leptonic branching fractions are
\begin{eqnarray*}
{\rm B}(\PW\ra\ev)&=&11.35\pm 0.46\stat\pm 0.17\syst\%,\\
{\rm B}(\PW\ra \mv)  &=& 11.10\pm 0.44\stat\pm 0.16\syst\%, \\
{\rm B}(\PW\ra \tv)  &=&  10.51\pm 0.55\stat\pm 0.22\syst\%,
\end{eqnarray*}
and are consistent with lepton universality and the Standard Model
expectations. 
Due to cross-contaminations in the identification of W decays to
\tv , \ev~or \mv , the measured ${\rm B}(\PW\ra \tv)$ is
26\% anticorrelated with ${\rm B}(\PW\ra\ev)$ and
24\% anticorrelated with ${\rm B}(\PW\ra\mv)$.
The ${\rm B}(\PW\ra\ev)$ is
4.6\% anticorrelated with ${\rm B}(\PW\ra\mv)$.

If lepton universality is assumed
a fit for
${\rm B}(\PW\ra \qqbar)$
and the total cross sections at the four energies
yields 

$$ {\rm B}(\PW\ra \qqbar) = 66.97 \pm 0.65 \stat\pm 0.32   \syst\%. $$ 
\noindent
This result can be expressed in terms of the individual couplings of the
W to quark-antiquark pairs:
\begin{eqnarray}
   \frac{{\rm B}(\PW\ra \qqbar)}{1-{\rm B}(\PW\ra \qqbar)} & = &
 ( |\ckm{ud}|^2 + |\ckm{cd}|^2 +
|\ckm{us}|^2 + |\ckm{cs}|^2 + |\ckm{ub}|^2 + |\ckm{cb}|^2 )
 (1+\alpha_{s}(\PMMW )/\pi).
\nonumber \\
\end{eqnarray}
The least well known of these is $|V_{\rm{cs}}|$. Using the world 
average value of $\alpha_{s}$(\PMMZ ) evolved to \PMMW~,
$\alpha_{s}$(\PMMW ) = $0.121 \pm 0.002$, and the squared sum of the other
measured CKM matrix elements~\cite{pdg98} which is 
$1.05 \pm 0.01$, the measured hadronic branching fraction is 
$$ |V_{\rm{cs}}| = 0.951\ \pm 0.030 \stat\pm 0.015
\syst. $$

\section{Conclusions}
The W-pair production cross section at $\sqrt{s}$ =  188.63~\GEV\
has been measured in all decay channels from an integrated
luminosity of 174.20~$\ipb$. The total cross section is found to be 
$$\PSW =  15.71 \pm 0.34\stat \pm 0.18\syst ~\pb. $$
\noindent
This result is 5.5\% (1.9 standard deviations) lower than 
the {\tt GENTLE} prediction but in better agreement with more recent 
calculations. It agrees well with the recent measurement by the
DELPHI Collaboration~\cite{delphi_xsec189} at the same CM energy. 

After inclusion of the data
taken at CM energies of 161, 172 and 183~\GEV, the
hadronic decay branching fraction is found to be $ 66.97 \pm 0.65
\stat \pm 0.32 \syst$\% which is used to determine the CKM matrix
element $|V_{\rm{cs}}|$ equal to $0.951 \pm 0.030\stat \pm 0.015\syst$.

\section*{Acknowledgements}
We would like to thank S.~Dittmaier and W.~Placzek for helpful
discussions on DPA. It is a pleasure to congratulate our
colleagues from the CERN accelerator divisions for the successful
operation of LEP at 189 GeV. We are indebted to the engineers and
technicians in all our institutions for their contributions to the
excellent performance of ALEPH. Those of us from non-member
countries thank CERN for its hospitality.

\clearpage


\section*{Appendix: Neural network input variables for the $\qqbar\qqbar$ 
selection}

The neural network hadronic event selection uses 14 variables.
These are based on
global event properties, heavy quark flavour tagging, jet properties and
\PWW\ kinematics and are listed below.
The four jets are numbered in order of decreasing energy.

\vspace{0.5cm}\noindent
{\em Global event properties}
\begin{itemize}
 \item Thrust
 \item Sphericity
 \item Missing energy
 \item Sum of the four smallest interjet angles
\end{itemize}
{\em Heavy flavour tagging}
\begin{itemize}
 \item Probability of an event being a light quark (uds) event based upon
impact parameter significance of charged particles in the event
\end{itemize}
{\em Jet properties}
\begin{itemize}
 \item Maximum electromagnetic energy fraction of a jet in any one degree cone
 \item Maximum summed charged particle energy fraction of a jet
 \item Minimum number of charged particles in a jet
\end{itemize}
{\em \PWW\ kinematics}
\begin{itemize}
 \item Angle between Jet2 and Jet3

The following jet related variables are determined from kinematically
fitted jet momenta.

 \item Energy of Jet1
 \item Energy of Jet3
 \item Energy of Jet4
 \item Smallest jet mass
 \item Second smallest jet mass
\end{itemize}


\end{document}

%% file: authb.tex
\pagestyle{empty}
\newpage
\small
%
%
\newlength{\saveparskip}
\newlength{\savetextheight}
\newlength{\savetopmargin}
\newlength{\savetextwidth}
\newlength{\saveoddsidemargin}
\newlength{\savetopsep}
\setlength{\saveparskip}{\parskip}
\setlength{\savetextheight}{\textheight}
\setlength{\savetopmargin}{\topmargin}
\setlength{\savetextwidth}{\textwidth}
\setlength{\saveoddsidemargin}{\oddsidemargin}
\setlength{\savetopsep}{\topsep}
%
%
\setlength{\parskip}{0.0cm}
\setlength{\textheight}{25.0cm}
\setlength{\topmargin}{-1.5cm}
\setlength{\textwidth}{16 cm}
\setlength{\oddsidemargin}{-0.0cm}
\setlength{\topsep}{1mm}
\pretolerance=10000
\centerline{\large\bf The ALEPH Collaboration}
\footnotesize
\vspace{0.5cm}
{\raggedbottom
\begin{sloppypar}
\samepage\noindent
R.~Barate,
D.~Decamp,
P.~Ghez,
C.~Goy,
S.~Jezequel,
J.-P.~Lees,
F.~Martin,
E.~Merle,
\mbox{M.-N.~Minard},
B.~Pietrzyk
\nopagebreak
\begin{center}
\parbox{15.5cm}{\sl\samepage
Laboratoire de Physique des Particules (LAPP), IN$^{2}$P$^{3}$-CNRS,
F-74019 Annecy-le-Vieux Cedex, France}
\end{center}\end{sloppypar}
\vspace{2mm}
\begin{sloppypar}
\noindent
R.~Alemany,
S.~Bravo,
M.P.~Casado,
M.~Chmeissani,
J.M.~Crespo,
E.~Fernandez,
M.~Fernandez-Bosman,
Ll.~Garrido,$^{15}$
E.~Graug\'{e}s,
M.~Martinez,
G.~Merino,
R.~Miquel,
Ll.M.~Mir,
A.~Pacheco,
H.~Ruiz
\nopagebreak
\begin{center}
\parbox{15.5cm}{\sl\samepage
Institut de F\'{i}sica d'Altes Energies, Universitat Aut\`{o}noma
de Barcelona, E-08193 Bellaterra (Barcelona), Spain$^{7}$}
\end{center}\end{sloppypar}
\vspace{2mm}
\begin{sloppypar}
\noindent
A.~Colaleo,
D.~Creanza,
M.~de~Palma,
G.~Iaselli,
G.~Maggi,
M.~Maggi,
S.~Nuzzo,
A.~Ranieri,
G.~Raso,
F.~Ruggieri,
G.~Selvaggi,
L.~Silvestris,
P.~Tempesta,
A.~Tricomi,$^{3}$
G.~Zito
\nopagebreak
\begin{center}
\parbox{15.5cm}{\sl\samepage
Dipartimento di Fisica, INFN Sezione di Bari, I-70126 Bari, Italy}
\end{center}\end{sloppypar}
\vspace{2mm}
\begin{sloppypar}
\noindent
X.~Huang,
J.~Lin,
Q. Ouyang,
T.~Wang,
Y.~Xie,
R.~Xu,
S.~Xue,
J.~Zhang,
L.~Zhang,
W.~Zhao
\nopagebreak
\begin{center}
\parbox{15.5cm}{\sl\samepage
Institute of High Energy Physics, Academia Sinica, Beijing, The People's
Republic of China$^{8}$}
\end{center}\end{sloppypar}
\vspace{2mm}
\begin{sloppypar}
\noindent
D.~Abbaneo,
G.~Boix,$^{6}$
O.~Buchm\"uller,
M.~Cattaneo,
F.~Cerutti,
G.~Dissertori,
H.~Drevermann,
R.W.~Forty,
M.~Frank,
F.~Gianotti,
T.C.~Greening,
A.W.~Halley,
J.B.~Hansen,
J.~Harvey,
P.~Janot,
B.~Jost,
M.~Kado,
V.~Lemaitre,
P.~Maley,
P.~Mato,
A.~Minten,
A.~Moutoussi,
F.~Ranjard,
L.~Rolandi,
D.~Schlatter,
M.~Schmitt,$^{20}$
O.~Schneider,$^{2}$
P.~Spagnolo,
W.~Tejessy,
F.~Teubert,
E.~Tournefier,
A.~Valassi,
J.J.~Ward,
A.E.~Wright
\nopagebreak
\begin{center}
\parbox{15.5cm}{\sl\samepage
European Laboratory for Particle Physics (CERN), CH-1211 Geneva 23,
Switzerland}
\end{center}\end{sloppypar}
\vspace{2mm}
\begin{sloppypar}
\noindent
Z.~Ajaltouni,
F.~Badaud,
G.~Chazelle,
O.~Deschamps,
S.~Dessagne,
A.~Falvard,
P.~Gay,
C.~Guicheney,
P.~Henrard,
J.~Jousset,
B.~Michel,
S.~Monteil,
\mbox{J-C.~Montret},
D.~Pallin,
J.M.~Pascolo,
P.~Perret,
F.~Podlyski
\nopagebreak
\begin{center}
\parbox{15.5cm}{\sl\samepage
Laboratoire de Physique Corpusculaire, Universit\'e Blaise Pascal,
IN$^{2}$P$^{3}$-CNRS, Clermont-Ferrand, F-63177 Aubi\`{e}re, France}
\end{center}\end{sloppypar}
\vspace{2mm}
\begin{sloppypar}
\noindent
J.D.~Hansen,
J.R.~Hansen,
P.H.~Hansen,$^{1}$
B.S.~Nilsson,
A.~W\"a\"an\"anen
\nopagebreak
\begin{center}
\parbox{15.5cm}{\sl\samepage
Niels Bohr Institute, 2100 Copenhagen, DK-Denmark$^{9}$}
\end{center}\end{sloppypar}
\vspace{2mm}
\begin{sloppypar}
\noindent
G.~Daskalakis,
A.~Kyriakis,
C.~Markou,
E.~Simopoulou,
A.~Vayaki
\nopagebreak
\begin{center}
\parbox{15.5cm}{\sl\samepage
Nuclear Research Center Demokritos (NRCD), GR-15310 Attiki, Greece}
\end{center}\end{sloppypar}
\vspace{2mm}
\begin{sloppypar}
\noindent
A.~Blondel,$^{12}$
\mbox{J.-C.~Brient},
F.~Machefert,
A.~Roug\'{e},
M.~Swynghedauw,
R.~Tanaka
\linebreak
H.~Videau
\nopagebreak
\begin{center}
\parbox{15.5cm}{\sl\samepage
Laboratoire de Physique Nucl\'eaire et des Hautes Energies, Ecole
Polytechnique, IN$^{2}$P$^{3}$-CNRS, \mbox{F-91128} Palaiseau Cedex, France}
\end{center}\end{sloppypar}
\vspace{2mm}
\begin{sloppypar}
\noindent
E.~Focardi,
G.~Parrini,
K.~Zachariadou
\nopagebreak
\begin{center}
\parbox{15.5cm}{\sl\samepage
Dipartimento di Fisica, Universit\`a di Firenze, INFN Sezione di Firenze,
I-50125 Firenze, Italy}
\end{center}\end{sloppypar}
\vspace{2mm}
\begin{sloppypar}
\noindent
A.~Antonelli,
G.~Bencivenni,
G.~Bologna,$^{4}$
F.~Bossi,
P.~Campana,
G.~Capon,
V.~Chiarella,
P.~Laurelli,
G.~Mannocchi,$^{1,5}$
F.~Murtas,
G.P.~Murtas,
L.~Passalacqua,
M.~Pepe-Altarelli
\nopagebreak
\begin{center}
\parbox{15.5cm}{\sl\samepage
Laboratori Nazionali dell'INFN (LNF-INFN), I-00044 Frascati, Italy}
\end{center}\end{sloppypar}
\vspace{2mm}
\begin{sloppypar}
\noindent
M.~Chalmers,
J.~Kennedy,
J.G.~Lynch,
P.~Negus,
V.~O'Shea,
B.~Raeven,
D.~Smith,
P.~Teixeira-Dias,
A.S.~Thompson
\nopagebreak
\begin{center}
\parbox{15.5cm}{\sl\samepage
Department of Physics and Astronomy, University of Glasgow, Glasgow G12
8QQ,United Kingdom$^{10}$}
\end{center}\end{sloppypar}
\begin{sloppypar}
\noindent
R.~Cavanaugh,
S.~Dhamotharan,
C.~Geweniger,$^{1}$
P.~Hanke,
V.~Hepp,
E.E.~Kluge,
G.~Leibenguth,
A.~Putzer,
K.~Tittel,
S.~Werner,$^{19}$
M.~Wunsch$^{19}$
\nopagebreak
\begin{center}
\parbox{15.5cm}{\sl\samepage
Kirchhoff-Institut f\"ur Physik, Universit\"at Heidelberg, D-69120
Heidelberg, Germany$^{16}$}
\end{center}\end{sloppypar}
\vspace{2mm}
\begin{sloppypar}
\noindent
R.~Beuselinck,
D.M.~Binnie,
W.~Cameron,
G.~Davies,
P.J.~Dornan,
M.~Girone,
N.~Marinelli,
J.~Nowell,
H.~Przysiezniak,$^{1}$
J.K.~Sedgbeer,
J.C.~Thompson,$^{14}$
E.~Thomson,$^{23}$
R.~White
\nopagebreak
\begin{center}
\parbox{15.5cm}{\sl\samepage
Department of Physics, Imperial College, London SW7 2BZ,
United Kingdom$^{10}$}
\end{center}\end{sloppypar}
\vspace{2mm}
\begin{sloppypar}
\noindent
V.M.~Ghete,
P.~Girtler,
E.~Kneringer,
D.~Kuhn,
G.~Rudolph
\nopagebreak
\begin{center}
\parbox{15.5cm}{\sl\samepage
Institut f\"ur Experimentalphysik, Universit\"at Innsbruck, A-6020
Innsbruck, Austria$^{18}$}
\end{center}\end{sloppypar}
\vspace{2mm}
\begin{sloppypar}
\noindent
C.K.~Bowdery,
P.G.~Buck,
D.P.~Clarke,
G.~Ellis,
A.J.~Finch,
F.~Foster,
G.~Hughes,
R.W.L.~Jones,
N.A.~Robertson,
M.~Smizanska
\nopagebreak
\begin{center}
\parbox{15.5cm}{\sl\samepage
Department of Physics, University of Lancaster, Lancaster LA1 4YB,
United Kingdom$^{10}$}
\end{center}\end{sloppypar}
\vspace{2mm}
\begin{sloppypar}
\noindent
I.~Giehl,
F.~H\"olldorfer,
K.~Jakobs,
K.~Kleinknecht,
M.~Kr\"ocker,
A.-S.~M\"uller,
H.-A.~N\"urnberger,
G.~Quast,$^{1}$
B.~Renk,
E.~Rohne,
H.-G.~Sander,
S.~Schmeling,
H.~Wachsmuth,
C.~Zeitnitz,
T.~Ziegler
\nopagebreak
\begin{center}
\parbox{15.5cm}{\sl\samepage
Institut f\"ur Physik, Universit\"at Mainz, D-55099 Mainz, Germany$^{16}$}
\end{center}\end{sloppypar}
\vspace{2mm}
\begin{sloppypar}
\noindent
A.~Bonissent,
J.~Carr,
P.~Coyle,
C.~Curtil,
A.~Ealet,
D.~Fouchez,
O.~Leroy,
T.~Kachelhoffer,
P.~Payre,
D.~Rousseau,
A.~Tilquin
\nopagebreak
\begin{center}
\parbox{15.5cm}{\sl\samepage
Centre de Physique des Particules de Marseille, Univ M\'editerran\'ee,
IN$^{2}$P$^{3}$-CNRS, F-13288 Marseille, France}
\end{center}\end{sloppypar}
\vspace{2mm}
\begin{sloppypar}
\noindent
M.~Aleppo,
M.~Antonelli,
S.~Gilardoni,
F.~Ragusa
\nopagebreak
\begin{center}
\parbox{15.5cm}{\sl\samepage
Dipartimento di Fisica, Universit\`a di Milano e INFN Sezione di
Milano, I-20133 Milano, Italy.}
\end{center}\end{sloppypar}
\vspace{2mm}
\begin{sloppypar}
\noindent
H.~Dietl,
G.~Ganis,
K.~H\"uttmann,
G.~L\"utjens,
C.~Mannert,
W.~M\"anner,
\mbox{H.-G.~Moser},
S.~Schael,
R.~Settles,$^{1}$
H.~Stenzel,
W.~Wiedenmann,
G.~Wolf
\nopagebreak
\begin{center}
\parbox{15.5cm}{\sl\samepage
Max-Planck-Institut f\"ur Physik, Werner-Heisenberg-Institut,
D-80805 M\"unchen, Germany\footnotemark[16]}
\end{center}\end{sloppypar}
\vspace{2mm}
\begin{sloppypar}
\noindent
P.~Azzurri,
J.~Boucrot,$^{1}$
O.~Callot,
M.~Davier,
L.~Duflot,
\mbox{J.-F.~Grivaz},
Ph.~Heusse,
A.~Jacholkowska,$^{1}$
L.~Serin,
\mbox{J.-J.~Veillet},
I.~Videau,$^{1}$
J.-B.~de~Vivie~de~R\'egie,
D.~Zerwas
\nopagebreak
\begin{center}
\parbox{15.5cm}{\sl\samepage
Laboratoire de l'Acc\'el\'erateur Lin\'eaire, Universit\'e de Paris-Sud,
IN$^{2}$P$^{3}$-CNRS, F-91898 Orsay Cedex, France}
\end{center}\end{sloppypar}
\vspace{2mm}
\begin{sloppypar}
\noindent
G.~Bagliesi,
T.~Boccali,
G.~Calderini,
V.~Ciulli,
L.~Fo\`a,
A.~Giassi,
F.~Ligabue,
A.~Messineo,
F.~Palla,$^{1}$
G.~Rizzo,
G.~Sanguinetti,
A.~Sciab\`a,
G.~Sguazzoni,
R.~Tenchini,$^{1}$
A.~Venturi,
P.G.~Verdini
\samepage
\begin{center}
\parbox{15.5cm}{\sl\samepage
Dipartimento di Fisica dell'Universit\`a, INFN Sezione di Pisa,
e Scuola Normale Superiore, I-56010 Pisa, Italy}
\end{center}\end{sloppypar}
\vspace{2mm}
\begin{sloppypar}
\noindent
G.A.~Blair,
J.~Coles,
G.~Cowan,
M.G.~Green,
D.E.~Hutchcroft,
L.T.~Jones,
T.~Medcalf,
J.A.~Strong,
\mbox{J.H.~von~Wimmersperg-Toeller}
\nopagebreak
\begin{center}
\parbox{15.5cm}{\sl\samepage
Department of Physics, Royal Holloway \& Bedford New College,
University of London, Surrey TW20 OEX, United Kingdom$^{10}$}
\end{center}\end{sloppypar}
\vspace{2mm}
\begin{sloppypar}
\noindent
R.W.~Clifft,
T.R.~Edgecock,
P.R.~Norton,
I.R.~Tomalin
\nopagebreak
\begin{center}
\parbox{15.5cm}{\sl\samepage
Particle Physics Dept., Rutherford Appleton Laboratory,
Chilton, Didcot, Oxon OX11 OQX, United Kingdom$^{10}$}
\end{center}\end{sloppypar}
\vspace{2mm}
\begin{sloppypar}
\noindent
\mbox{B.~Bloch-Devaux},
P.~Colas,
B.~Fabbro,
G.~Fa\"if,
E.~Lan\c{c}on,
\mbox{M.-C.~Lemaire},
E.~Locci,
P.~Perez,
J.~Rander,
\mbox{J.-F.~Renardy},
A.~Rosowsky,
P.~Seager,$^{13}$
A.~Trabelsi,$^{21}$
B.~Tuchming,
B.~Vallage
\nopagebreak
\begin{center}
\parbox{15.5cm}{\sl\samepage
CEA, DAPNIA/Service de Physique des Particules,
CE-Saclay, F-91191 Gif-sur-Yvette Cedex, France$^{17}$}
\end{center}\end{sloppypar}
\vspace{2mm}
\begin{sloppypar}
\noindent
S.N.~Black,
J.H.~Dann,
C.~Loomis,
H.Y.~Kim,
N.~Konstantinidis,
A.M.~Litke,
M.A. McNeil,
G.~Taylor
\nopagebreak
\begin{center}
\parbox{15.5cm}{\sl\samepage
Institute for Particle Physics, University of California at
Santa Cruz, Santa Cruz, CA 95064, USA$^{22}$}
\end{center}\end{sloppypar}
\vspace{2mm}
\begin{sloppypar}
\noindent
C.N.~Booth,
S.~Cartwright,
F.~Combley,
P.N.~Hodgson,
M.~Lehto,
L.F.~Thompson
\nopagebreak
\begin{center}
\parbox{15.5cm}{\sl\samepage
Department of Physics, University of Sheffield, Sheffield S3 7RH,
United Kingdom$^{10}$}
\end{center}\end{sloppypar}
\vspace{2mm}
\begin{sloppypar}
\noindent
K.~Affholderbach,
A.~B\"ohrer,
S.~Brandt,
C.~Grupen,
J.~Hess,
A.~Misiejuk,
G.~Prange,
U.~Sieler
\nopagebreak
\begin{center}
\parbox{15.5cm}{\sl\samepage
Fachbereich Physik, Universit\"at Siegen, D-57068 Siegen, Germany$^{16}$}
\end{center}\end{sloppypar}
\vspace{2mm}
\begin{sloppypar}
\noindent
C.~Borean,
G.~Giannini,
B.~Gobbo
\nopagebreak
\begin{center}
\parbox{15.5cm}{\sl\samepage
Dipartimento di Fisica, Universit\`a di Trieste e INFN Sezione di Trieste,
I-34127 Trieste, Italy}
\end{center}\end{sloppypar}
\vspace{2mm}
\begin{sloppypar}
\noindent
H.~He,
J.~Putz,
J.~Rothberg,
S.~Wasserbaech
\nopagebreak
\begin{center}
\parbox{15.5cm}{\sl\samepage
Experimental Elementary Particle Physics, University of Washington, WA 98195
Seattle, U.S.A.}
\end{center}\end{sloppypar}
\vspace{2mm}
\begin{sloppypar}
\noindent
S.R.~Armstrong,
K.~Cranmer,
P.~Elmer,
D.P.S.~Ferguson,
Y.~Gao,
S.~Gonz\'{a}lez,
O.J.~Hayes,
H.~Hu,
S.~Jin,
J.~Kile,
P.A.~McNamara III,
J.~Nielsen,
W.~Orejudos,
Y.B.~Pan,
Y.~Saadi,
I.J.~Scott,
J.~Walsh,
J.~Wu,
Sau~Lan~Wu,
X.~Wu,
G.~Zobernig
\nopagebreak
\begin{center}
\parbox{15.5cm}{\sl\samepage
Department of Physics, University of Wisconsin, Madison, WI 53706,
USA$^{11}$}
\end{center}\end{sloppypar}
}
\footnotetext[1]{Also at CERN, 1211 Geneva 23, Switzerland.}
\footnotetext[2]{Now at Universit\'e de Lausanne, 1015 Lausanne, Switzerland.}
\footnotetext[3]{Also at Dipartimento di Fisica di Catania and INFN Sezione di
 Catania, 95129 Catania, Italy.}
\footnotetext[4]{Also Istituto di Fisica Generale, Universit\`{a} di
Torino, 10125 Torino, Italy.}
\footnotetext[5]{Also Istituto di Cosmo-Geofisica del C.N.R., Torino,
Italy.}
\footnotetext[6]{Supported by the Commission of the European Communities,
contract ERBFMBICT982894.}
\footnotetext[7]{Supported by CICYT, Spain.}
\footnotetext[8]{Supported by the National Science Foundation of China.}
\footnotetext[9]{Supported by the Danish Natural Science Research Council.}
\footnotetext[10]{Supported by the UK Particle Physics and Astronomy Research
Council.}
\footnotetext[11]{Supported by the US Department of Energy, grant
DE-FG0295-ER40896.}
\footnotetext[12]{Now at Departement de Physique Corpusculaire, Universit\'e de
Gen\`eve, 1211 Gen\`eve 4, Switzerland.}
\footnotetext[13]{Supported by the Commission of the European Communities,
contract ERBFMBICT982874.}
\footnotetext[14]{Also at Rutherford Appleton Laboratory, Chilton, Didcot, UK.}
\footnotetext[15]{Permanent address: Universitat de Barcelona, 08208 Barcelona,
Spain.}
\footnotetext[16]{Supported by the Bundesministerium f\"ur Bildung,
Wissenschaft, Forschung und Technologie, Germany.}
\footnotetext[17]{Supported by the Direction des Sciences de la
Mati\`ere, C.E.A.}
\footnotetext[18]{Supported by the Austrian Ministry for Science and Transport.}
\footnotetext[19]{Now at SAP AG, 69185 Walldorf, Germany}
\footnotetext[20]{Now at Harvard University, Cambridge, MA 02138, U.S.A.}
\footnotetext[21]{Now at D\'epartement de Physique, Facult\'e des Sciences de Tunis, 1060 Le Belv\'ed\`ere, Tunisia.}
\footnotetext[22]{Supported by the US Department of Energy,
grant DE-FG03-92ER40689.}
\footnotetext[23]{Now at Department of Physics, Ohio State University, Columbus, OH 43210-1106, U.S.A.}
%
\setlength{\parskip}{\saveparskip}
\setlength{\textheight}{\savetextheight}
\setlength{\topmargin}{\savetopmargin}
\setlength{\textwidth}{\savetextwidth}
\setlength{\oddsidemargin}{\saveoddsidemargin}
\setlength{\topsep}{\savetopsep}
\normalsize
\newpage
\pagestyle{plain}
\setcounter{page}{1}

%% file: wwxsec189_final.bbl
\begin{thebibliography}{99}

\bibitem{xsecA}
ALEPH Collaboration,
{\it Measurement of the W mass in ${\mathrm e^+ e^-}$
collisions at production threshold},
~\PL{401}{1997}{347};
{\it Measurement of W-pair cross section in \Pepem\
collisions at 172 GeV},
~\PL{415}{1997}{435};
{\it Measurement of W-pair production in \Pepem\ collisions at 183GeV},
~\PL{453}{1999}{107}.

\bibitem{det} ALEPH Collaboration,
{\it ALEPH: A detector for electron-positron annihilations at LEP},
~\NIM{A 294}{1990}{121}.

\bibitem{perf} ALEPH Collaboration,
{\it Performance of the ALEPH detector at LEP},
~\NIM{A 360}{1995}{481}.

\bibitem{ALEPHlcal}  ALEPH Collaboration,
{\it Measurement of the absolute luminosity with the ALEPH detector},
~Z. Phys. {\bf C53} (1992) 375.

\bibitem{BHLUMI}
S.~Jadach et al.,
~\PL{253}{1991}{469};~\PL{257}{1991}{173};~\PL{260}{1991}{438};
Comp. Phys. Commun. {\bf 70} (1992) 305.
Also: S.~Jadach, M.~Melles, B.F.L. Ward and S.A.~Yost,
UTHEP-98-050.

\bibitem{LEPbeam}
The LEP Energy Working group,
{\it Evaluation of the
 LEP Centre-of-Mass Energy for Data Taken in 1998},
LEP ECAL/99-01.

\bibitem{LEP2workshop}
W.~Beenakker and F.A.~Berends, in Physics at LEP2, CERN 96-01,
eds. G.~Altarelli, T.~Sj\"ostrand and F.~Zwirner, Vol. 1, p. 79.

\bibitem{GENTLE}
D. Bardin et al., Nucl.\ Phys.\ (Proc.\ Suppl.) {\bf B37} (1994) 148;
D.~Bardin et al.,
Comp. Phys. Commun. {\bf 104} (1997) 161.

\bibitem{KORALW}
 M.~Skrzypek, S.~Jadach, W.~Placzek and Z.~W\c{a}s,
 Comp. Phys. Commun. {\bf 94} (1996) 216.

\bibitem{Jetset}
 T.~Sj\"ostrand, Comp. Phys. Commun. {\bf 82} (1994) 74.

\bibitem{EXCALIBUR}
 F.A.~Berends, R.~Pittau and R.~Kleiss,
  Comp. Phys. Commun.  {\bf 85} (1995) 437.

\bibitem{grc4f}
J.~Fujimoto et al., Comp. Phys. Commun. {\bf 100} (1997) 128.


\bibitem{KORALZ}
 S.~Jadach, B.F.L.~Ward and Z.~W\c{a}s, Comp. Phys. Commun. {\bf  79} (1994) 
503.

\bibitem{PYTHIA}
 T.~Sj\"ostrand, Comp. Phys. Commun. {\bf 82} (1994) 74.

\bibitem{PHOT02}
 ALEPH Collaboration,
{\it An experimental study of $\gamma\gamma\ra$ hadrons at LEP},
~\PL{313}{1993}{509}.

\bibitem{BHWIDE}
S.~Jadach et al., \PL{390}{1997}{298}.

\bibitem{UNIBAB}
 H.~Anlauf et al., Comp. Phys. Commun.  {\bf 79} (1994) 466.

\bibitem{durham}
Yu.L. Dokshitzer, J. Phys. {\bf G17} (1991) 1441.

\bibitem{wmass_189}
ALEPH Collaboration,
{\it Measurement of the W mass and width in ${\mathrm e^+ e^-}$ collisions at
189~\GEV},
~to be published in Eur. Phys. J. C

\bibitem{HERWIG}
G.~Marchesini et al., Comp. Phys. Commun.  {\bf 67} (1992) 465.

\bibitem{BEJetset}
L.~L\"onnblad and T.~Sj\"ostrand,
Eur. Phys. J. {\bf C2} (1998) 165.



\bibitem{Beenakker}
        W.~Beenakker, F.A.~Berends and A.P.~Chapovsky,
        Nucl.\ Phys.\ {\bf B548} (1999) 3.

\bibitem{YFSWW}
        S.~Jadach et al.,
        Phys.\ Lett.\ {\bf B417} (1998) 326; CERN-TH 99-222, UTHEP 98-0502, 
        submitted to Phys. ReV. D. 

\bibitem{RacoonWW}
        A.~Denner, S.~Dittmaier, M.~Roth and D.~Wackeroth,
        {\em ``$\mathcal{O}\rm(\alpha)$ corrections to
        $\mathrm{e}^+ \mathrm{e}^- \rightarrow \mathrm{WW} \rightarrow
        \mathrm{4\ fermions}(+\gamma)$:
        first numerical results from RacoonWW''},
        BI-TP~99/45, Dec. 1999, hep-ph/9912261.


\bibitem{MCLEP2}
        S.~Dittmaier, private communication and LEP2 Monte-Carlo Workshop, 
        March 2000.

\bibitem{MWWA}
The LEP and SLD Collaborations, {\it A Combination of Preliminary
Electroweak Measurements and Constraints on the Standard Model},
CERN-EP-2000-016, Jan. 2000.

\bibitem{pdg98}
C.~Caso et al.\ (Particle Data Group), Review of Particles
Physics, Eur. Phys. J. {\bf C3} (1998) 1.

\bibitem{delphi_xsec189}
DELPHI Collaboration,
{\it W pair production cross-section and branching fractions in 
${\mathrm e^+ e^-}$ interactions at 189~\GEV},
~CERN-EP-2000-035.

\end{thebibliography}
